# Highly sensitive NO$_2$ sensors by pulsed laser deposition on graphene

Margus Kodu, Artjom Berholts, Tauno Kahro, Tea Avarmaa, Aarne Kasikov, Ahti Niilisk, Harry Alles, and Raivo Jaaniso[a]

*Institute of Physics, University of Tartu, W. Ostwald Street 1, EE-50411 Tartu, Estonia*

**Abstract**

Graphene as a single-atomic-layer material is fully exposed to environmental factors and has therefore a great potential for the creation of sensitive gas sensors. However, in order to realize this potential for different polluting gases, graphene has to be functionalized - adsorption centers of different types and with high affinity to target gases have to be created at its surface. In the present work, the modification of graphene by small amounts of laser-ablated materials is introduced for this purpose as a versatile and precise tool. The approach has been demonstrated with two very different materials chosen for pulsed laser deposition (PLD) – a metal (Ag) and a dielectric oxide (ZrO$_2$). It was shown that the gas response and its recovery rate can be significantly enhanced by choosing the PLD target material and deposition conditions. The response to NO$_2$ gas in air was amplified up to 40 times in the case of PLD-modified graphene, in comparison with pristine graphene, and it reached 7-8% at 40 ppb of NO$_2$ and 20-30% at 1 ppm of NO$_2$. The PLD process was conducted in a background gas (5 x10$^{-2}$ mbar oxygen or nitrogen) and resulted in the atomic areal densities of the deposited materials of about 10$^{15}$ cm$^{-2}$. The ultimate level of NO$_2$ detection in air, as extrapolated from the experimental data obtained at room temperature under mild ultraviolet excitation, was below 1 ppb.

[a] Author to whom correspondence should be addressed. Electronic mail: raivo.jaaniso@ut.ee

Graphene, a one-atom-thick carbon material, is considered as one of the main building blocks for the next generation electronic and sensor applications.[1] The full exposure of graphene surface to environment, together with easy doping[2] and low electrical noise of this material,[3] paves the way for the fabrication of sensitive chemical sensors. For example, the detection of $NO_2$ gas on ppq-level[4] or even single adsorbed molecules[3,5] has been demonstrated in high vacuum or in inert (Ar) gas atmosphere. As the binding energy between the $NO_2$ molecule and the surface of pristine graphene is small (<0.07 eV),[6] the functionalization of graphene by internal defects (vacancies, their complexes, edges, grain boundaries) or by foreign dopants, adatoms, clusters, or molecular groups has been proposed for increasing the gas sensitivity.[7,8] In this way not only the binding energy (and therefore the surface coverage) can be increased but also the electronic properties of graphene can be modified so that the adsorption of a gas molecule results in an increased electrical response. Another challenge for creating the full-value sensors with high sensitivity is their ability to work not only in inert atmosphere but also in air.

$NO_2$ is one of the key atmospheric pollutant gases, the content of which in outdoor air is monitored in the environment and is regulated worldwide.[9,10] The semiconductor sensors have been developed[11] which are capable of sensing $NO_2$ in air at low concentrations, however, these devices are power-consuming because of the relatively high working temperature. Sub-ppm-level $NO_2$ sensing was recently demonstrated at low working temperatures by nanopatterning of graphene[12] or by deposition of metal layers on it.[13]

In this work, we are demonstrating the functionalization of a single-layer (>95%) CVD graphene by pulsed laser deposition (PLD) for an enhanced sensing of $NO_2$ gas in air. In order to show the versatility of the technique, two materials belonging to different classes - a metal (Ag) and an oxide ($ZrO_2$) - were chosen as PLD targets. These concrete materials were chosen because of their different work functions $\Phi$ with respect to graphene ($\Phi(Ag)<\Phi(graphene)<\Phi(ZrO_2)$; see analysis below), which was expected to lead to different (n- or p-) type of doping from these materials and hence to different change of sensor signals.

The graphene was grown in a hot-wall quartz tube CVD reactor and thereafter transferred from copper foils onto substrates, 10×10 $mm^2$ $Si/SiO_2$ with a 300 nm thick thermal oxide layer, as described before.[14] The size of the graphene sheet grown on the copper foil was 15×23 $mm^2$; smaller pieces (≈5×6 $mm^2$) were cut from it for transferring onto the substrates. The Ti(3 nm)/Au(60 nm) electrodes were made on the substrates by electron beam deposition through a shadow mask before the graphene transfer. In the PLD process, either ceramic $ZrO_2$ or metallic Ag pellet was ablated by a focused beam of a KrF excimer laser. Before starting the PLD procedure, the PLD chamber was evacuated to $10^{-6}$ mbar and the sensor substrates were heated in-situ at 150 °C for 1.5 hours and then cooled down to room temperature. This procedure did not induce any defects in graphene as verified by Raman spectroscopy (see Fig. S1 in the supplementary material). The $ZrO_2$ was ablated by using a laser pulse energy density of 2.5 $J/cm^2$ on the target and $5 \times 10^{-2}$ mbar oxygen pressure in the chamber. For the deposition of Ag, the laser energy density was 5 $J/cm^2$ and the nitrogen gas at $5 \times 10^{-2}$ mbar was used as the background. The laser fluences were chosen so that the deposition rates of two materials were approximately equal. More details about the PLD method have been described previously.[15] The measurements of electrical characteristics and gas sensitivity were carried out in a small (7 $cm^3$) test chamber by Keithley 2400 SourceMeter. The total gas flow through the sample chamber was kept constant at 200 sccm and the flow rates of individual gases ($N_2$, $O_2$, and $NO_2/N_2$ mixture; all 99.999% pure) were varied by mass flow controllers (Brooks, model SLA5820) for changing the $NO_2$ content in the background of synthetic air. It is known that the illumination with UV light can enhance the response of graphene gas sensors.[4,14] In our experiments, we used for illumination a Xe-Hg lamp (L2422, Hamamatsu) with band-pass (365 nm) filter.

The graphene samples on electrode substrates were modified in a sequence of PLD processes, which transferred a small amount of target material onto the sample, whereby the amount of the deposited material in each process was precisely controlled by the number of laser pulses. The sample scheme is shown in Fig. S2 of the supplementary material. After each deposition, the sample was removed from the PLD chamber and its characteristics (Raman spectrum, conductance, gas response) were measured. Then the next deposition was made according to the procedure described above. Fig. 1 shows the Raman spectra taken with a Renishaw inVia spectrometer at 514.5 nm excitation after different depositions, whereby the cumulative number of PLD laser pulses *N* is written at each curve.

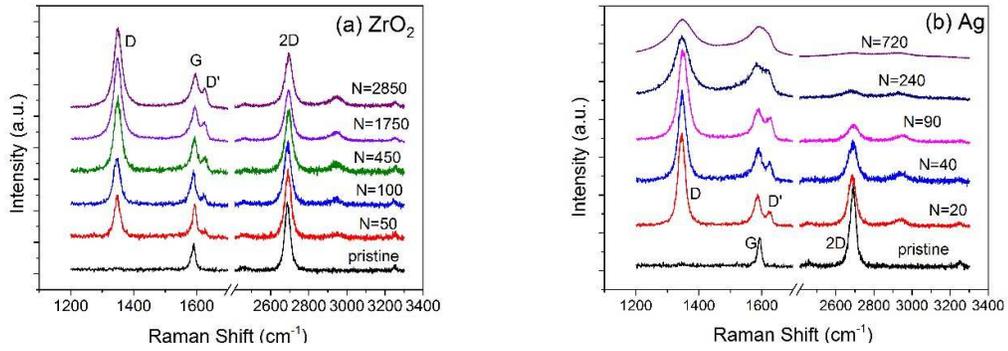

Fig.1. Raman spectra of graphene sensor samples with pulsed-laser-deposited $ZrO_2$ (a) and Ag (b) at different cumulative numbers *N* of the laser pulses used in the processes.

For comparison, the Raman spectra of pristine graphene samples are given at the bottom. The G- and 2D-bands were located at ≈1586 cm$^{-1}$ and ≈2686 cm$^{-1}$ and their full widths at half maximums were ≈15 cm$^{-1}$ and ≈30 cm$^{-1}$, respectively. In all samples, the defect-related D-band was initially absent in the Raman spectra, but it appeared and rapidly increased in size already at small amounts of the deposited material. The appearance of an intensive D-peak at about 1350 cm$^{-1}$ as well as a D'-peak at about 1625 cm$^{-1}$ indicates that the PLD process has created defects in graphene.[16] The broadening of Raman bands occurred to a different extent in the case of two target materials. In the case of $ZrO_2$ the broadening of D-band became saturated at about 29 cm$^{-1}$ level after 450 laser pulses, whereas in the case of Ag it continued up to 720 pulses, when it reached 120 cm$^{-1}$, indicating a rather substantial disordering of graphene.

Fig. 2 presents the scanning electron microscopy (SEM; Helios NanoLab 600) images of the graphene sensors after deposition of 2850 pulses of $ZrO_2$ (a) and 720 pulses of Ag (b). The SEM images were taken only after the last deposition in order to avoid the effect of electron beam irradiation on gas sensitivity in the interim measurements.

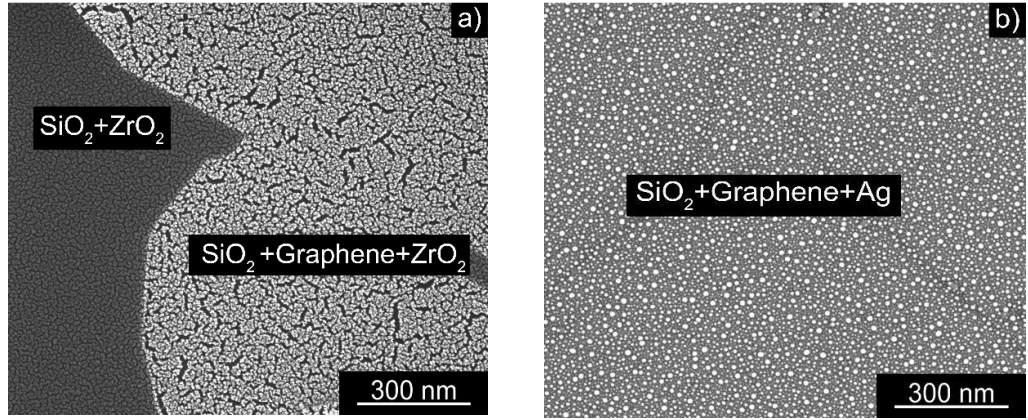

Fig.2. SEM images of graphene sensors with pulsed laser-deposited $ZrO_2$ (a) and Ag (b).

As one can see, the surface morphologies of the two samples are very different – $ZrO_2$ shows a porous granular structure, whereas Ag is gathered into small clusters. It becomes even more evident if one adds the data (see Table I) obtained at this stage from spectrometric ellipsometry (GES-5E) and x-ray fluorescence (XRF; Rigaku ZSX-400). The ellipsometric data show that in the case of $ZrO_2$ as a deposited material the pores and voids fill about half of the coating, but in the case of Ag the nanoclusters cover only about 1/10 of the surface. It also follows from Table I that the thickness values obtained by two different techniques ($d_1$ and $d_2$) do coincide at least within ±10% for a given material.

Table I. Thickness and density parameters of deposited layers.

| Material | $d_E$ (nm) | PP (%) | $d_1$ (nm) | m (µg/cm$^2$) | $d_2$ (nm) | N | n (cm$^{-2}$) |
|---|---|---|---|---|---|---|---|
| $ZrO_2$ | 14 | 47 | 6.6 | 3.55 | 6.5 | 2850 | 6.1×10$^{12}$ |
| Ag | 10 | 89 | 1.1 | 0.96 | 0.9 | 720 | 7.4×10$^{12}$ |

$d_E$ – nominal thickness of the deposited layer, determined from ellipsometric data

PP – percentage of pores or empty spaces as determined from ellipsometric data

$d_1=d_E\times(100-PP)/100$ – thickness of the deposited material without pores

m – areal mass density as determined by XRF

$d_2=m/\rho$ – thickness of the deposited material without pores ($\rho$ – density of material; $\rho(ZrO_2)$=5.49 g/cm$^3$; $\rho(Ag)$=10.5 g/cm$^3$)

N – total number of laser pulses used at deposition

n – areal atomic density of the deposited material per laser pulse

The sensor characteristics were initially recorded on pristine samples and then after each PLD procedure. The results of gas response measurements, recorded at room temperature for the two deposited materials, are shown in Fig. 3. The gas response is defined as the relative change of conductance. On each subfigure, four different cycles are shown, where the pure synthetic air was interchanged with the synthetic air containing $NO_2$ at different concentrations from 40 ppb to 1 ppm.

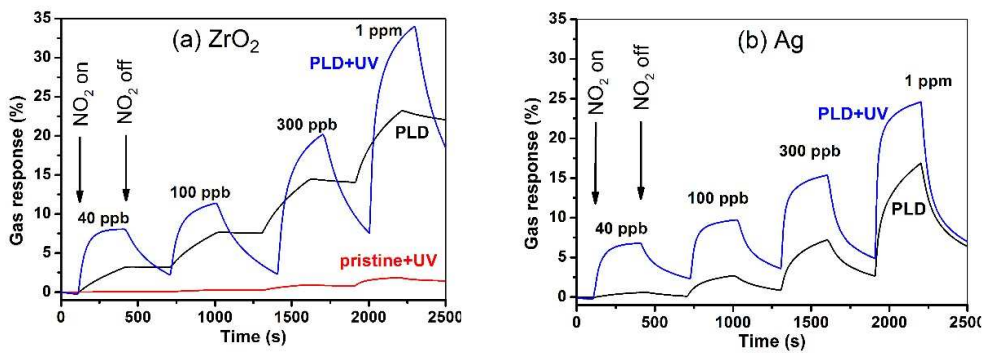

Fig. 3. Temporal responses to $NO_2$ gas in the dark and under UV illumination at different gas concentrations for pristine and $ZrO_2$-modified (a; after 2850 pulses) and Ag-modified (b; after 135 pulses) graphene. The wavelength of UV light was 365 nm and its intensity 20 mW/cm$^2$. Responses were recorded in dry air at room temperature (22 °C).

One can see large enhancements in sensor responses for PLD-modified samples. In fact, two different responses are shown for these samples – with and without UV illumination of the sensor surface. The illumination with a light of moderate intensity (20 mW/cm$^2$) further improved the gas response as well as its temporal characteristics. Note that for a given slit area (4 mm$^2$), the power of light required for reversible operation of the sensor at room temperature is below 1 mW and it can be many orders smaller in downscaled devices. In the case of pristine samples, the response to $NO_2$ gas was systematically observable only under UV illumination, being 2-4% at $NO_2$ concentration of 1 ppm (see Fig. 3a).

The response was amplified in PLD-modified samples by about 10 times at 1 ppm and even more, by about 40 times, at the lowest gas concentration (40 ppb) probed. There was also a clear difference between the two materials used for the sensitization of graphene sensors. First, the response speed was faster in the case of Ag, as compared to $ZrO_2$, and this comparison was valid with and without UV illumination. Second, without UV illumination the recovery of the signal in pure air was almost absent in the case of $ZrO_2$, but a (partial) recovery of the signal occurred in the case of Ag. The response times, determined by fitting the temporal curves with the exponential function, were improved from 500 s to 50 s in the case of $ZrO_2$ by applying UV excitation. In the case of Ag, the temporal response could be fitted with the double exponential function, which yielded characteristic times 12 s and 112 s in the case of UV excitation at $NO_2$ concentration of 1 ppm. The response speed was 2-3 times slower in dark conditions.

In Fig. 4a, the $NO_2$ gas response of graphene sensors is shown as a function of the areal density of the deposited material on graphene. The numbers of laser pulses used in depositions were converted into the areal densities of deposits by using the coefficients given in the last column of Table I. The response was defined here as a relative change of conductance, measured at 300 s after the step change of $NO_2$ gas concentration from 0 to 1 ppm in synthetic air. As one can see, the gas responses increase quickly to about 20% and then attain a nearly constant value at the areal atomic densities between $0.5 \times 10^{15}$ and $2 \times 10^{15}$ cm$^{-2}$. At larger amounts of deposits, the curves in Fig. 4a split – in the case of $ZrO_2$ the sensor response remains roughly constant (at least up to areal density $1.7 \times 10^{16}$ cm$^{-2}$, which corresponds to the maximum number of laser pulses $N$=2850; data point not shown in the figure), but there is a significant drop of sensitivity in the case of Ag. This is probably related to a significantly faster decrease of the overall conductance in the case of Ag (Fig. 4b).

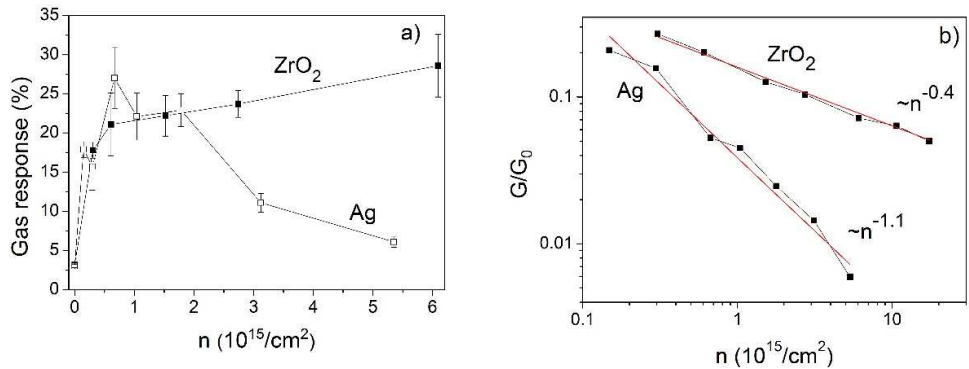

Fig. 4. The dependence of the sensor's $NO_2$ gas response (a) and normalized conductance (b) on the areal density of the deposited materials on graphene surface.

It follows from Fig. 4b that the conductance $G$ of the samples measured in air at room temperature decreased steadily with the increasing amount of the deposited material. This decrease followed the power law with power indices -0.4 and -1.1 for $ZrO_2$ and Ag, respectively. A stronger influence of Ag as compared to $ZrO_2$ can also be seen from the evaluation of the Raman spectra in Fig. 1. There is a significant broadening of all Raman bands and almost a complete disappearance of 2D band at the highest number of laser pulses in the case of Ag deposition.

There are several possible reasons for the difference of Raman spectra and conductance in the case of the two deposited materials. First, it may be that the kinetic energies of ablated atoms are higher in the case of Ag, because of the higher laser energy density used at its ablation. This could lead to a higher impact energy of ablated atoms (ions) with graphene, consequently, to a higher probability of defect formation, and finally, to a more significant scattering of charge carriers. Note that other conditions, including the gas environment in the deposition chamber (same pressure, just a small difference in the stopping power of $O_2$ and $N_2$ gases), were practically similar in the case of both materials. Another reason may be the difference in work functions. As the work function of Ag (4.3 eV) is smaller than that of graphene (4.7 eV), the doping by Ag clusters leads to a decrease in the initial p-type conductivity. The work function of $ZrO_2$ can be derived from the following data: the electron affinity $\chi$=2.5-3 eV and bandgap energy $E_g$=5.7-5.8 eV.[17,18] Pure $ZrO_2$ has a p-type conductivity at normal conditions,[19] which means that the work function is larger than $\chi+E_g/2$>5.35 eV. Consequently, its work function is bigger as compared to graphene and a doping from $ZrO_2$ will increase the p-type conduction of graphene. At the same time, one should take into account the decreasing mobility with the increasing number of defects in the case of both materials, the net effect being a faster decrease in conduction in the case of Ag, as compared to $ZrO_2$. Finally, morphology may play a role when the amount of the deposited material increases. According to Fig. 1 and the data of Table I, the screening of graphene by a previously deposited material is more significant in the case of $ZrO_2$, which probably contributes to a relatively slow decrease of conductance and finally stops the influence of the deposited material on graphene's Raman spectra.

In the case of Ag nanoparticles attention has to be paid to Ag oxidation state. The binding energy of Ag $3d_{5/2}$ level (367.9 eV) was determined from X-ray photoelectron spectrum and the energy of $M_4N_{45}N_{45}$ transition (358.2 eV) from Auger spectrum. The sum of these values - Auger parameter - is close to the value assigned to Ag(0) and is clearly different from the values assigned to Ag in oxides.[20] There may be a minor contribution of $Ag_2O$ on the surface of silver nanoparticles because of oxidizing conditions (atmospheric ozone or sub-ppm $NO_2$)[21] but it will not change the interpretation given above because the oxide can be present only in a very small amount and because its work function (4.8-4.9 eV)[22] is also smaller than that of $ZrO_2$.

The density of the induced defects $n_D$ can be estimated from the ratio $I(D)/I(G)$ of the integral intensities of D- and G-bands in the Raman spectra. It has been shown that this dependence has a quite general character with a maximum of the ratio (equal to about 3 in the case of 514 nm excitation) occurring at $n_D \approx 8 \times 10^{12}$ cm$^{-1}$.[23] The maximum value of I(D)/I(G) was obtained after the first (Ag) and the third ($ZrO_2$) deposition, when the areal densities of the deposited material were $1.5 \times 10^{14}$ cm$^{-1}$ and $2.7×10^{15}$, respectively. Consequently, only a small part of deposited atoms or ions – 5% in the case of Ag and 0.3% in the case of $ZrO_2$ - were active in forming the defects altering the Raman spectra. This can be explained by the fact that the laser-ablated species have a wide distribution of kinetic energies and only the high-energy tail of this distribution exceeds the defect creation threshold.

In the PLD process, the amount of high-energy particles can easily be regulated. The initial state of the ablated material can be described by a "shifted" Maxwellian distribution with a characteristic temperature of $≈10^4$ K, which starts moving towards the substrate with a speed of $≈10^4$ m/s.[24] Consequently, the average energies of ablated species are of the order of 100 eV and in the case of PLD in high vacuum the energy of most atoms or ions at the substrate surpass the threshold knock-on energy for ejecting an in-lattice carbon atom (17 eV[25]). However, when the plume propagates in the background gas, as in the present work, the propagation speed and plume temperature will decrease as a result of multiple collisions with gas molecules. Eventually, the plume will be completely thermalized when the pressure of the background gas $P >> 0.1$ mbar at typical target-to-substrate distances of several centimeters.[24]

In our case, the gas pressure is in the intermediate range ($P=5·10^{-2}$ mbar) and a small part of atoms or ions hitting the graphene still has enough energy for creating defects. The energy distribution depends not only on the gas pressure, but also on the laser interaction with the target material (and, in case of ns-pulses, with the initial plasma).[24] Larger laser fluences were used in the case of Ag target for obtaining a similar deposition rate as in the case of $ZrO_2$ target (see the last column in Table I), but, at the same time, the larger laser fluence seemingly resulted in a bigger number of high-energy particles.

The enhancement of gas response after PLD can be explained both by the increase of the density of adsorption sites and, at the same time, bigger affinity of these sites towards the target gas, especially under UV exposure. First principle studies have shown, indeed, that $NO_2$ adsorption on regular graphene planes is relatively weak at room temperature (adsorption energy $E_a < 0.07$ eV[6]), but it can be easily adsorbed on defective or doped graphene ($E_a =0.3-3$ eV[26,27]). In a PLD process the doping may have a multistage character – the energetic edge of the ablated material creates vacancies, which can then be filled by less energetic particles "smoothly landing" onto graphene. Such way of doping is technologically simpler than the forming of vacancies by PLD and adding dopants by electron beam deposition or sputtering.[28]

The PLD-induced adsorption centers can be the point defects formed by atomic (ionic) bombardment, i.e. the same defects which influence the Raman spectra and conductivity. However, as $n_D \ll n$, the adsorption centers may have a different origin as well. One possibility is the phase boundary between graphene and the edges of the clusters formed by the deposited material.[13]

The effect of UV light has previously been ascribed to the cleaning of graphene surface and, hence, to liberating the active sites for target gas adsorption.[4,14] It follows from Fig. 3 that the illumination preferentially activates the adsorption sites with a larger affinity towards $NO_2$. In addition to significantly increased sensitivity at lower concentrations, it also induces a faster recovery of the signal via $NO_2$ desorption in the case of $ZrO_2$. In dark conditions the desorption of $NO_2$ was almost absent for this material, but occurred partially for Ag deposit. Clearly, the adsorption sites with a very different origin, electronic behavior and binding energies are involved in these two cases.

To conclude, we demonstrated that pulsed laser deposition is a versatile and precise tool for sensitizing graphene-based gas sensors. Versatility means that a variety of defects or clusters can be formed on graphene from different atomic species with different "landing energies", and preciseness refers to the fact that typically only 1% or less of a monolayer is deposited by a single laser pulse. Our studies showed that the gas response can be significantly enhanced after creating adsorption centers by PLD. The response and recovery times of gas response strongly depend on the type of the PLD target material. The signal recovery times and sensitivity at low concentrations could be further improved by illumination with a modest-intensity UV light. The ultimate level of detection, as extrapolated from the experimental data, was below 1 ppb. Hence, the practical range of the measurement for the environmental monitoring of $NO_2$ (from a few to a few hundreds of ppbs) can be well covered with the sensors based on PLD-modified graphene.

See supplementary material for the sample scheme and additional Raman data.

The research leading to these results has received funding from the European Union's Horizon 2020 research and innovation program under grant agreement No 696656, and from the Estonian Research Council by institutional grants IUT34-27 and IUT2-24. The authors are grateful to Peeter Ritslaid for measuring XRF data, Leonard Matisen for measuring XPS/AES data and Jelena Kozlova for taking SEM images.